# Sub-nm wide electron channels protected by topology

Christian Pauly[1], Bertold Rasche[2], Klaus Koepernik[3], Marcus Liebmann[1], Marco Pratzer[1], Manuel Richter[3], Jens Kellner[1], Markus Eschbach[4], Bernhard Kaufmann[1], Lukasz Plucinski[4], Claus M. Schneider[4], Michael Ruck[2,5], Jeroen van den Brink[3], and Markus Morgenstern[1]

[1]II. Institute of Physics B and JARA-FIT, RWTH Aachen University, D-52074 Aachen, Germany

[2]Department of Chemistry and Food Chemistry, TU Dresden, D-01062 Dresden, Germany

[3]Leibniz Institute for Solid State and Materials Research, IFW Dresden, P.O. box 270116, D-01171 Dresden, Germany and Dresden Center for Computational Science

[4]Peter Grünberg Institute PGI-6, Forschungszentrum Jülich, D-52425 Jülich, Germany

[5]Max Planck Institute for Chemical Physics of Solids, D-01187 Dresden, Germany

**Helical locking of spin and momentum and prohibited backscattering are the key properties of topologically protected states.[1-2] They are expected to enable novel types of information processing such as spintronics[3] by providing pure spin currents[4], or fault tolerant quantum computation by using the Majorana fermions at interfaces of topological states with superconductors[5]. So far, the required helical conduction channels used to realize Majorana fermions are generated through application of an axial magnetic field to conventional semiconductor nanowires[6]. Avoiding the magnetic field enhances the possibilities for circuit design significantly[7]. Here, we show that sub-nanometer wide electron channels with natural helicity are present at surface step-edges of the recently discovered topological insulator $Bi_{14}Rh_3I_9$ [8]. Scanning tunneling spectroscopy reveals the electron channels to be continuous in both energy and space within a large band gap of 200 meV, thereby, evidencing its non-trivial topology. The absence of these channels in the closely related, but topologically trivial insulator $Bi_{13}Pt_3I_7$ corroborates the channels' topological nature. The back-scatter-free electron channels are a direct consequence of $Bi_{14}Rh_3I_9$'s structure, a stack of 2D topologically insulating, graphene-like planes separated by trivial insulators. We demonstrate that the surface of $Bi_{14}Rh_3I_9$ can be engraved using an atomic force microscope, allowing networks of protected channels to be patterned with nm precision.**

The compound $Bi_{14}Rh_3I_9$ consists of two types of layers being alternately stacked. One layer, $[(Bi_4Rh)_3I]^{2+}$, exhibits a graphene-like honeycomb lattice formed by rhodium-centered bismuth cubes as revealed by X-ray diffraction (XRD) (red layer, Fig. 1b) and is a 2D topological insulator (2DTI) according to density functional theory (DFT)[8]. Its structure mimics the originally proposed quantum spin Hall insulator in graphene[9], but with an inverted band gap being four orders of magnitude larger. The other layer separating the 2DTIs is a $[Bi_2I_8]^{2-}$ spacer with a trivial band gap (blue layer, Fig. 1b). Such a stack of layers has been proposed to be a weak 3D topological insulator (3DTI)[10], but remained elusive until DFT confirmed the synthesized compound $Bi_{14}Rh_3I_9$ to be one[8]. Theory predicts that weak 3DTIs feature helical edge states at each step edge of the surface, which is perpendicular to the stacking direction[11]. These edge states are topologically protected and immune to backscattering as long as time-reversal symmetry persists. Thus, perfect conduction of these channels with conductivity $e^2/h$ is anticipated[11,12]. Moreover, partially interfacing these channels with superconductors is predicted to induce Majorana fermions at the rim of the interfacial region.[5]

Scanning tunneling microscopy (STM) on the cleaved surface of $Bi_{14}Rh_3I_9$, i.e. the surface perpendicular to the stacking direction, identifies the spacer layer and the 2DTI layer by their different atomic-scale structures (Fig. 1c and 1d) and their corresponding different step heights (inset, Fig. 1a). The graphene-like honeycomb lattice (Fig. 1c) belongs to the 2DTI layer and exhibits the unit cell size known from XRD[8], whereas the hexagonally arranged spots in the other layer (Fig. 1d) fit to the iodide ions in the spacer layer $[Bi_2I_8]^{2-}$. Combining the STM step height of both layers (1.2 nm) provides a good agreement with the respective layer thickness deduced from XRD (1.25 nm)[8]. The 2DTI step edges are mostly zigzag-terminated (Fig. 1e, inset) with a few kinks and a few adsorbates on top (Fig. 1e), the latter probably due to remaining iodide ions from the spacer layer (methods).

Figure 1f shows the locally measured differential conductivity $dI/dV(V)$, known to represent the local density of states (LDOS)[13], as recorded on the 2DTI layer (red curve), on the spacer layer (blue curve) and at the step edge (grey curve). On the 2DTI layer, we find a gap between $V = -180$ mV and $V = -360$ mV (remaining intensity within this gap is explained in methods). This is in excellent agreement with the gap measured by ARPES[8] at -170 to -370 meV below the Fermi level $E_F$. Importantly, there is strong dI/dV intensity within this 2DTI gap if measured at the step edge. This indicates the edge state. The peak maximum is at the lower part of the band gap in accordance with the dispersion from tight-binding calculations[14]. A larger gap is revealed on the spacer layer as expected. These properties are present on all areas of the sample, partly with a different intensity distribution, which is attributed to different local chemistry or to a different density of states of the probing tip. Figure 1h shows a spatially resolved *dI/dV* map corresponding to the topography of Fig. 1g and measured at a sample voltage within the band gap of the 2DTI. Bright stripes at all step edges indicate the presence of an edge mode, as also found on all other step edges of the 2DTI layer.

Another example of the edge state is shown in Fig. 2a, where its Bloch type character appears as an oscillation with unit cell periodicity along the zigzag direction (Fig. 2c). A profile line across the edge state exhibits a full width at half maximum (FWHM) of 0.83 nm only (Fig. 2b). This is an upper limit due to possible convolution effects with the tip shape. Thus, the edge state is confined to a single unit cell (width: 0.92 nm[8]) as predicted by tight-binding calculations[14]. Such a width is much smaller than for edge states of the buried 2DTI made of HgTe quantum wells (edge state width: ~200 nm)[12,15] implying the possibility of much smaller devices. Theory predicts that the helical conduction remains robust for step edge heights containing any number of exposed stacks and is even stabilized by disorder.[16-20] Thus, simply scratching the surface deeper than a single layer induces a one-dimensional electron channel with a robust conductivity of at minimum $e^2/h$.[18] The first step in that direction is shown in Fig. 2d. Scratches about 3 layers deep are produced by atomic force microscopy (AFM). The distances between the centres of scratches are well below 100 nm and the edge channels are partly separated by 25 nm only. In order to make such networks operational, it is mandatory to move the non-trivial band gap to $E_F$. This may be achieved by surface doping (e.g. by Iodine) recalling that the calculated bulk position of $E_F$ is already within the non-trivial band gap[8]. DFT slab calculations further reveal n-type doping of the cleaved surface as observed experimentally.[8]

Next, we probe, if the edge state covers the whole non-trivial band gap as predicted by topology. Figure 2e is a colour code representation of the energy dependent LDOS across the step edge. It reveals pronounced edge state intensity in the whole non-trivial band gap and slightly weaker intensity even in the energy region below. This characteristic has been found at all ten probed step edges with lengths from 6 nm to 40 nm. Figure 3a, b shows another example of a 2DTI step edge again with an extremely narrow edge state running along the edge and being pushed around kinks (Fig. 3a). The edge state intensity is visible along the whole edge at all energies within the band gap (Fig. 3b, see also Supplementary Video) and again weaker even at energies below (-400 mV, -450 mV).

It, moreover, shows that intensity fluctuations of the edge state along the edge are small (except at energies below the gap) as expected for the prohibited backscattering. In order to sustain this assumption, the electron wave length as deduced from tight binding calculations[14] $\lambda_{\text{tight-binding}}$ is added to Fig. 2a and 3a. Obviously, there is no structure with periodicity $\lambda_{\text{tight-binding}}/2$, i.e. no standing electron waves, in remarkable contrast to conventional 1D electron systems, where such oscillations exhibit LDOS intensity oscillations close to 100 % [21].

Notice that edge states barely prone to backscattering have also been observed on some of the step edges of the 2DTI Bi bilayers on Bi(111)[22] or $Bi_2Te_3(0001)$[23], but in both cases the edge states energetically overlap with bulk states.

The topology of the edge state additionally requires its ubiquity at step edges. Figure 3c shows a case study for a rather disordered step edge. The intensity at the step edge strongly fluctuates, but the *dI/dV(V)* spectra always reveal fingerprints of the edge state. This becomes clear when zooming in on different regions near the step edge (Fig. 3c (i)-(v)). The spectra on the 2DTI layer in region (i) are flat within the band gap region with slightly different intensities in different regions due to a varying coupling to the underlying stripes of the spacer layer as deduced from comparison with DFT calculations (see section S1 of the Supplementary Information). Comparing these spectra with the spectra at most parts of the step edge, for instance region (iii), shows a strong peak indicating the edge state. The sharpness of the peak depends on details of the tip, e.g., it appears sharper in region (ii) which is measured after a tip switch. Zooming into an area of small edge state intensity, region (iv), located around a kink position (see topography in section S1 of the Supplementary Information), reveals the same peak, but a factor of 10 lower in intensity indicating a smaller coupling of the edge state to the tunnelling tip. Moreover, the edge state intensity is pushed to the right, i.e., it moves around the obstacle as predicted by Ringel *et al.*[16]. The size of the peak is even further reduced below a pair of adsorbates (region (v)) but can still be identified around $V = -270$ mV (see arrow). Concluding, we find signatures of a spatially continuous edge state within all investigated step edge areas, pointing to a robust character with respect to disorder as expected from a topologically protected state.

To further consolidate the topological character of the edge states in $Bi_{14}Rh_3I_9$, we have investigated the very similar system $Bi_{13}Pt_3I_7$, where Rh is replaced by the heavier Pt. The chemical composition is slightly different such that every second spacer layer is replaced by a single layer of iodide ions[24] (see structural model in Fig. 4e). The honeycomb layer (Fig. 4e) is again a 2DTI as revealed by DFT, whereas the compound $Bi_{13}Pt_3I_7$ itself is semi-metallic and topologically trivial (DFT results are depicted in section S2 of the Supplementary Information). The two different spacers lead to an alternating coupling between adjacent 2DTI layers, giving rise to a "dimerization", which, according to theory, is a unique possibility to render a weak 3DTI topologically trivial[16-20], such that the topologically protected edge states disappear.

STM images of the cleaved surface of $Bi_{13}Pt_3I_7$ (Fig. 4a) again exhibit two different layers, one with hexagonally arranged spots (Fig. 4c), triangularly reconstructed, identified as the insulating $[Bi_2I_8]^{2-}$ spacer and the other with a honeycomb structure (Fig. 4d) rendering it the 2DTI. The 2DTI exhibits single atoms on top, most probably remaining iodide ions from the spacers. Interestingly, such iodide ions are absent within the last two unit cells close to the zigzag step edges (Fig. 4h, topography). Pure iodide layers (the other spacer) have not been observed and most step heights (Fig. 4f) cover a complete unit cell of the dimerized layers in stacking direction (2.1 nm). The band structure of $Bi_{13}Pt_3I_7$ as measured by ARPES at photon energy $hv = 21.2$ eV (Fig. 4g) reveals two band openings around the Γ-point. These gaps are found to be trivial in DFT calculations albeit originating from non-trivial gaps of the 2DTI (section S2 of the Supplementary Information). In agreement, STS does not show any edge states within these band gaps (Fig. 4h, i) on all ten step edges probed. Notice that the contrast in Fig. 4h is chosen identical to Fig. 3b, where the edge state of $Bi_{14}Rh_3I_9$ is clearly apparent. Thus, the "dimerized" structure of $Bi_{13}Pt_3I_7$, where stacks are built from pairs of 2DTIs, is a trivial insulator without protected edge states as predicted by topological analysis. [16-20]

The new type of helical edge states in $Bi_{14}Rh_3I_9$ offers the opportunity to design spin filters[4] with extremely small footprint compared to 2DTIs in heterostructures[12]. Moreover, the interfacing with other materials such as superconductors or magnetic insulators required for advanced quantum circuitry[5,7] becomes directly accessible by shadow mask evaporation. In this sense, the discovery of the first weak 3DTI $Bi_{14}Rh_3I_9$ might offer similar advantages as graphene does with respect to conventional semiconductor heterostructures[25].

## Methods:

**Preparation and growth of $Bi_{14}Rh_3I_9$ crystals**

$Bi_{14}Rh_3I_9$ crystals were grown by thermal annealing of a stoichiometric mixture of Bi, Rh, and $BiI_3$ (molar ratio 11: 3: 3). The starting materials were ground under argon atmosphere in a glovebox. The homogeneous powder was sealed in an evacuated silica ampule and heated to 700 °C in a tubular furnace at a rate of approximately 600 K/h. Fast cooling to 420 °C (−4 K/min) and then slow cooling to 365 °C (−1 K/h) followed instantaneously. After three days the ampule was quenched in water. A more detailed description can be found in the work of Rasche *et al.*[27].

**Preparation and growth of $Bi_{13}Pt_3I_7$ crystals**

$Bi_{13}Pt_3I_7$ crystals were grown by thermal annealing of a stoichiometric mixture of Bi, Pt, and $BiI_3$ (molar ratio 10.67: 3: 2.33). The starting materials were ground under argon atmosphere in a glovebox. The homogeneous powder was sealed in an evacuated silica ampule and heated to 650 °C in a tubular furnace at a rate of approximately 600 K/h. Fast cooling to 500 °C (−4 K/min) and then slow cooling to 380 °C (−1 K/h) followed instantaneously. After three days the ampule was quenched in water.

**Preparation of atomically clean surfaces in ultrahigh vacuum (UHV)**

The $Bi_{14}Rh_3I_9$ and $Bi_{13}Pt_3I_7$ crystals (~ 1 mm in diameter) are mounted to a Mo sample holder by a carbon conductive adhesive. Prior to the STM and ARPES measurements, the samples were cleaved in-situ at room temperature using a commercial copper tape, leading to atomically flat terraces on the surface as checked by AFM and STM. The clean sample has been immediately transferred into the microscope (STM) or manipulator (ARPES) and cooled down to 6 K or 15 K, respectively. We checked that the remaining adsorbates at the step edges (Fig. 1e) are not originating from the background pressure. Therefore, we varied the background pressure during cleavage and transfer to the STM by a factor of ten. Furthermore, we varied the time between cleavage and transfer into the cryostat, which exhibits an excellent cryogenic vacuum, also by a factor of ten. Both processes did not change the density of adsorbates at the step edges. We therefore conclude that these adsorbates arise from the cleavage process itself and originate most probably from the spacer layers.

**STM and STS measurements**

The microscopic measurements have been performed in a home-built scanning tunneling microscope in UHV ($p_{base}$ = 10$^{-11}$ mbar) at $T$ = 6 K[28]. Topographic STM images and *dI/dV* images are recorded in constant-current mode with voltage *V* applied to the sample. The *dI/dV(V)* spectra are recorded after stabilizing the tip at a sample voltage $V_{stab}$ and a current $I_{stab}$ before opening the feedback loop. All spectroscopic measurements are carried out using a lock-in technique with a modulation frequency $\nu$ = 1.4 kHz and amplitude $V_{mod}$ = 4 mV resulting in an energy resolution $\delta E = \sqrt{(3.3 k_B T)^2 + (1.8 e V_{mod})^2} \approx 7$ meV (*e*: electron charge, $k_B$: Boltzmann constant)[29]. To first order, *dI/dV(V)* is proportional to the local density of states (LDOS) of the actual sample position at energy *E* with respect to the Fermi level $E_{F,sample}$ of the sample ($E = E_{F,sample} + eV$). It is also proportional to the LDOS of the tip at the Fermi level $E_{F,tip}$ of the tip. However, if the tip LDOS at $E_{F,sample}$ is not vanishing, one gets an additional contribution from the LDOS of the tip at energy *E* according to $E = E_{F,tip} - eV$ multiplied by the LDOS of the sample at $E_{F,sample}$[29]. Since $E_{F,sample}$ is crossing a bulk band within the 2DTI layer, the later contribution is also present, if the voltage refers to a band gap of the sample. This explains the remaining *dI/dV* intensity at voltages corresponding to the 2DTI band gap. The apparent pseudo-gap at $E_{F,sample}$ does not cure this contribution due to the finite energy resolution of the experiment[29].

**AFM experiments**

The scratches are produced within a commercial AFM (Bruker) using a carbon coated silicon cantilever in AFM contact mode at ambient conditions (contact force during scratching: $F = 10^{-6}$ N). The AFM images have been recorded in the tapping mode using the same carbon coated silicon cantilever (resonance frequency 275.1 kHz, force constant 43 N/m, oscillation amplitude 30 nm, set point 70%, velocity 2 μm/s).

**ARPES measurements**

ARPES spectra were measured in UHV ($p_{base} = 6·10^{-11}$ mbar) on the cleaved $Bi_{13}Pt_3I_7$ samples kept at 15 K using He I ($hv = 21.2$ eV) discharge within a laboratory based system. The overall energy resolution is 10 meV and the angular resolution is 0.6°. The spectra were recorded within 60 min after cleavage. The fact that the beam spot of the incident light is of the order of the sample size (~ 1 mm in diameter) causes some background intensity in the ARPES data probably originating from the carbon conductive adhesive. For the same reason, a slight softening of the bands in the spectra is observable.

**Computational details**

All band structure calculations were performed with the Full-Potential Local-Orbital (FPLO) code[30] version 14.00, within the local density approximation (LDA) using the parameterization PW92 according to Perdew et al.[31]. The Blöchl corrected linear tetrahedron method with a 8×8×4 k-mesh for $Bi_{13}Pt_3I_7$ and a 12x12x1 k-mesh for the single 2DTI layer $[Bi_{12}Pt_3I]^{3+}$ was employed. Spin-orbit coupling is treated on the level of the four-component Dirac-equation. The following basis states are treated as valence states: Bi:5s, 5p, 5d, 6s, 7s, 6p, 7p, 6d; Pt: 5s, 5p, 5d, 6s, 6p, 6d, 7s; I: 4s, 4p, 4d, 5s, 6s, 5p, 6p, 5d.

For the band structure calculation of $Bi_{13}Pt_3I_7$, we used the atomic structure as deduced from XRD experiments[24]. The single 2DTI layer $[Bi_{12}Pt_3I]^{3+}$ was calculated with an iodide layer placed above and beneath the $[Bi_{12}Pt_3I]^{3+}$ layer in order to adjust the charge. 10 Å of vacuum was added in $c^*$-direction in order to separate adjacent layers in a 3D supercell geometry. The cell constants and atomic positions were optimized within the *p6/mmm* layer symmetry.

The calculation of the topological invariants was implemented following[32] using the fact that the crystals all have an inversion symmetry such that parities of the states at the time-reversal invariant momenta (TRIMs) can be used to calculate the four $Z_2$ indices $v_0;(v_1v_2v_3)$[10,32-34]. More details can be found in the Supplementary Information of Rasche *et al.*[8]

## **Acknowledgements:**

We acknowledge financial support by the German science foundation via Mo 858/13-1, IS 250/1-1 and RI 932/7-1 being part of the Priority Programme "Topological Insulators" (SPP 1666). C. P. thanks the Fonds National de la Recherche (Luxembourg) for funding. We acknowledge helpful discussions with H. Obuse and C. Mudry as well as careful reading of the manuscript of R. McNeil.

**Author contributions:**

C.P. carried out the STM measurements and prepared the first version of the manuscript together with M.M. Material synthesis and XRD analysis were done by B.R. under the supervision of M. Ruck. J.K., M.E., and L.P. performed and analyzed the ARPES measurements under the supervision of C.M.S. Band structures have been calculated by B.R., K.K., and M. Richter under the supervision of J.v.d.B with K.K. implementing the calculation of topological invariants. B.K. produced the scratching patterns by AFM. All authors contributed to the data analysis, the general outline of the manuscript and its improvement. M.M. provided the general idea of the experiment and coordinated the project together with C.P..

Correspondence and requests for materials should be addressed to M.M. (mmorgens@physik.rwth-aachen.de).

**Figure captions:**

**Figure 1: Identification of the edge state. a,** STM image of cleaved $Bi_{14}Rh_3I_9$ ($V = 1$ V, $I = 100$ pA). Inset: height profile along the green line. **b,** Atomic polyhedron model of $Bi_{14}Rh_3I_9$ as deduced from XRD[8]. **c, d,** Zoom into the marked areas of (**a**) displaying the two different layers. Atomic model structure is overlaid (color code as in (**b**)). ((**c**) $V = 1.5$ V, $I = 100$ pA, (**d**) $V = -1.3$ V, $I = 100$ pA). **e,** STM image with step edge of a 2DTI layer ($V = 0.8$ V, $I = 100$ pA). Inset: zoom into the step edge region with overlaid honeycomb lattice revealing the zigzag termination (dotted lines are guides to the eye). **f,** $dI/dV(V)$ spectra taken at the positions marked in (**e**) by respectively colored squares ($V_{stab} = 0.8$ V, $I_{stab} = 100$ pA, $V_{mod} = 4$ mV). Notice the linearly vanishing $dI/dV$ intensity around $E_F$ which is attributed to a 2D Coulomb gap of Efros-Shklovskii type[26]. **g,** STM and **h,** $dI/dV$ image within the 2DTI band gap of a region with step edges of the 2DTI layer ($V = -250$ mV, $I = 100$ pA, $V_{mod} = 4$ mV). Strong $dI/dV$ intensity appears at the step edges.

**Figure 2**: **Edge state properties. a,** $dI/dV$ image within the bulk band gap ($V_{stab} = -337$ mV, $I_{stab} = 100$ pA, $V_{mod} = 4$ mV) of the step edge area shown in the inset as a STM image ($V = 0.8$ V, $I = 100$ pA). Rectangles mark the areas of profile lines in (**b**) and (**c**). The double arrow marks the electron wave length of the edge state at this particular energy as deduced from tight-binding calculations[14]. **b,** Profile line perpendicular to the step edge and averaged in the parallel direction over the blue rectangle in (**a**) with FWHM of the edge state marked. **c,** Profile line along the step edge taken from the olive rectangle with marked peak distance corresponding to the size of one unit cell. **d,** AFM image of the $Bi_{14}Rh_3I_9$ surface after scratching a network of step edges into the surface by a carbon coated Si cantilever. Profile line along the blue line is shown in the inset. **e,** Color plot of

*dI/dV(V)* spectra taken across the step edge. Three different lateral regions are separated by dotted lines and labeled within the topographic profile below. Different energetic features are marked.

**Figure 3: Non-trivial character of the edge state: continuity in energy and space. a,** STM image (top, $V$ = -262 mV, $I$ = 100 pA) and *dI/dV* image within the band gap (bottom, $V_{stab}$ = -262 mV, $I_{stab}$ = 100 pA, $V_{mod}$ = 4 mV) of a rather straight step edge. A kink position is marked. The double arrow marks the expected electron wave length at this particular energy as deduced from tight-binding calculations[14]. **b,** Stacked *dI/dV* images ($I_{stab}$ = 100 pA, $V_{mod}$ = 4 mV) of the area shown in **(a)** recorded at voltages across the band gap as marked on the left. **c,** *dI/dV* image (averaged from $V$ = -180 mV to -350 mV, $I_{stab}$ = 80 pA, $V_{mod}$ = 4 mV) including an undulated step edge, and *dI/dV(V)* spectra ($V_{stab}$ = 0.8 V, $I_{stab}$ = 80 pA, $V_{mod}$ = 4 mV) colored with respect to their area of origin marked in the *dI/dV* image. 2DTI layer and spacer layer are labeled and surrounded by dotted lines. The spectra in (i) originate from the 2DTI layer, the grey spectra in (ii)-(v) from the step edge region, the blue and red curves in (ii)-(v) from the spacer and the 2DTI layer, respectively. Shaded areas in (i)-(v) mark the bulk band gap as deduced from ARPES.

**Figure 4**: **Absence of a topological edge state in $Bi_{13}Pt_3I_7$. a,** STM image ($V$ = 1 V, $I$ = 100 pA) of cleaved $Bi_{13}Pt_3I_7$. **b,** Height profile along green line. **c, d,** Atomically resolved images of the two different layers as marked ((**c**) $V$ = 1 V, $I$ = 100 pA, (**d**) $V$ = 0.6 V, $I$ = 100 pA). **e,** Atomic model of $Bi_{13}Pt_3I_7$ as deduced from XRD[24] with different layers marked. **f,** STM image ($V$ = 1 V, $I$ = 100 pA) showing step heights of one unit cell (2.1 nm). **g,** ARPES intensity plot ($hv$ = 21.2 eV); full lines with double arrows mark two band gaps, dotted lines the energies of *dI/dV* plots in (**h**). **h,**

Stacked *dI/dV* images ($I_{stab}$ = 100 pA, $V_{mod}$ = 8 mV) of the area shown in the background STM image (*V* = 0.6 V, *I* = 100 pA) and recorded within the band gaps at voltages marked on the left, same contrast as in Fig. 3b. **i,** Local *dI/dV(V)* spectra ($V_{stab}$ = 1 V, $I_{stab}$ = 100 pA, $V_{mod}$ = 8 mV) recorded at the positions marked by arrows in (**h**) and on the insulating spacer layer. Band gaps deduced from ARPES are marked in red.

**Figures:**

**Fig.1**

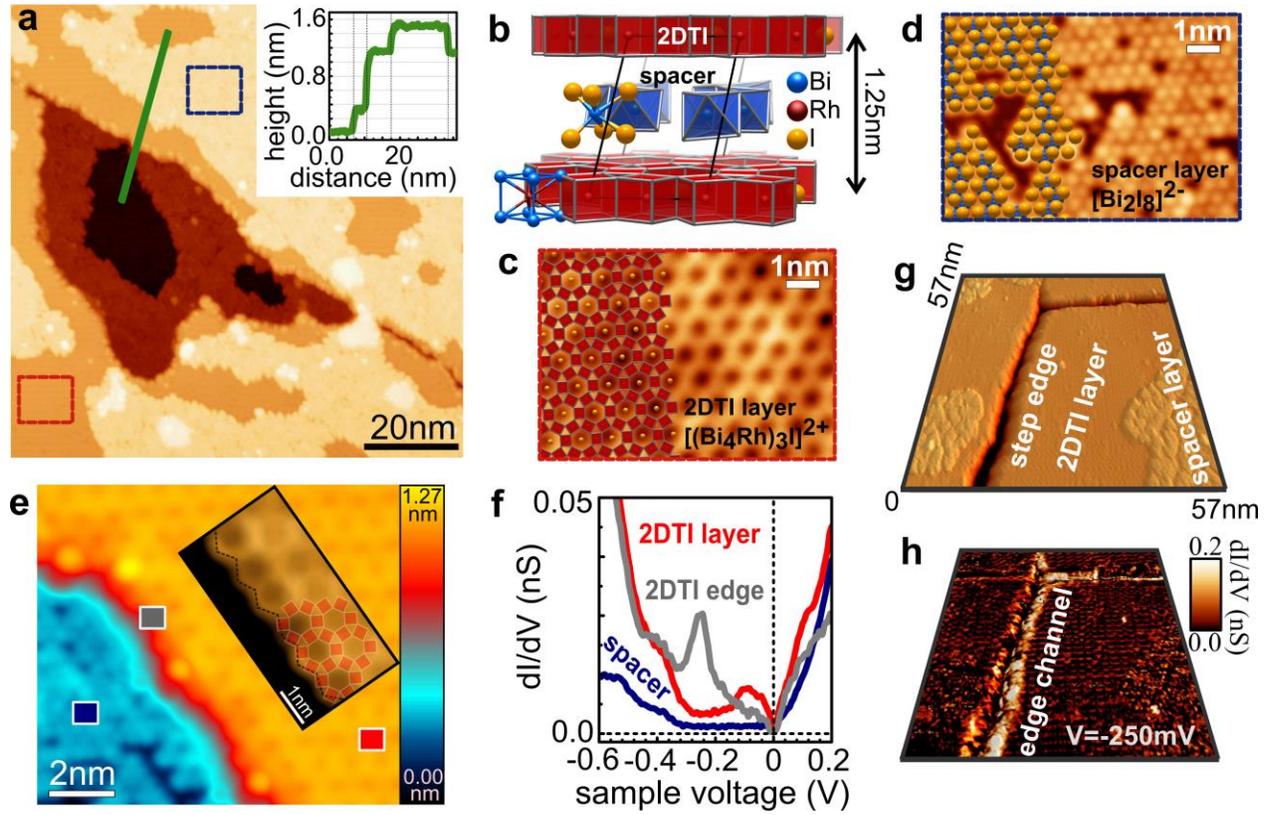

**Fig.2**

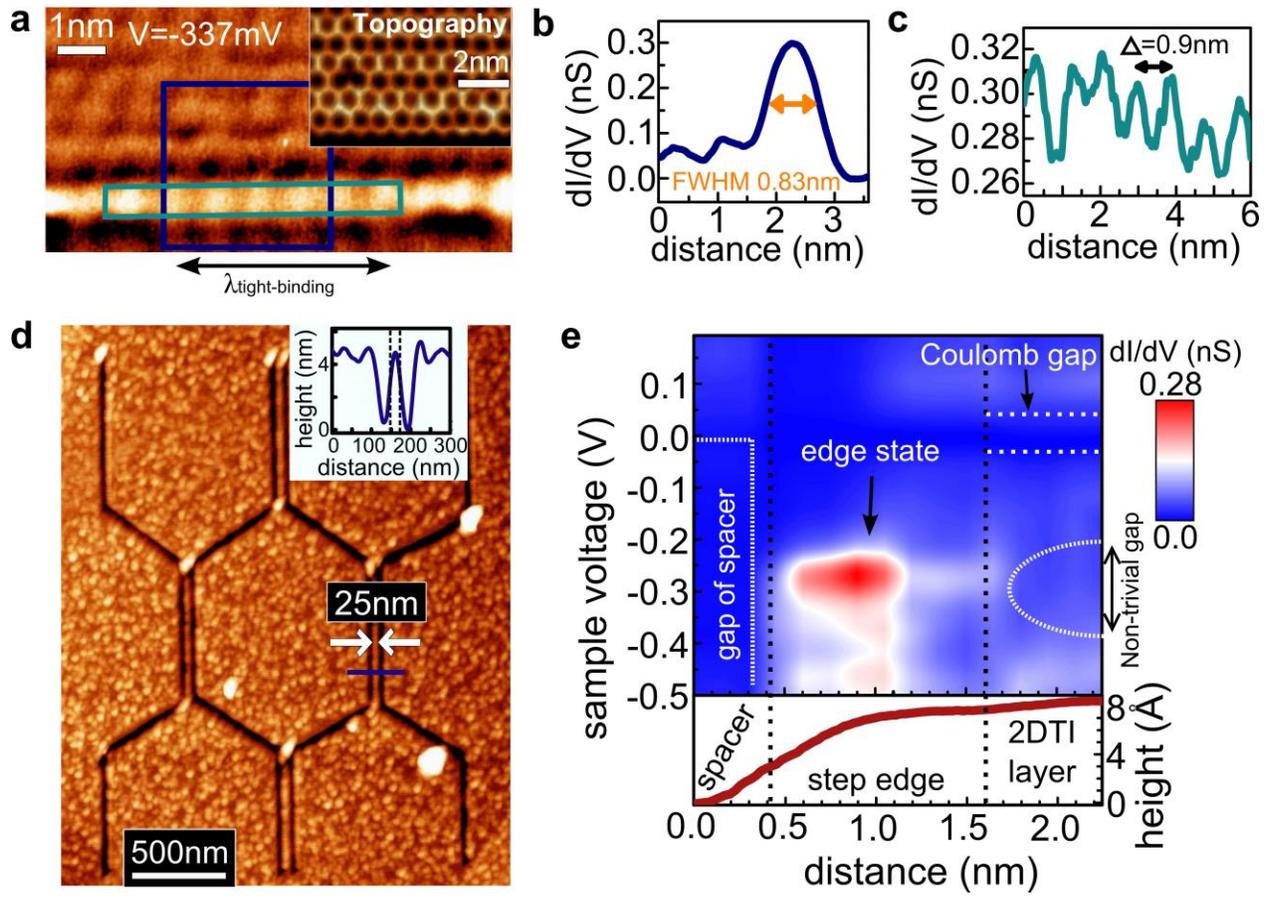

**Fig.3**

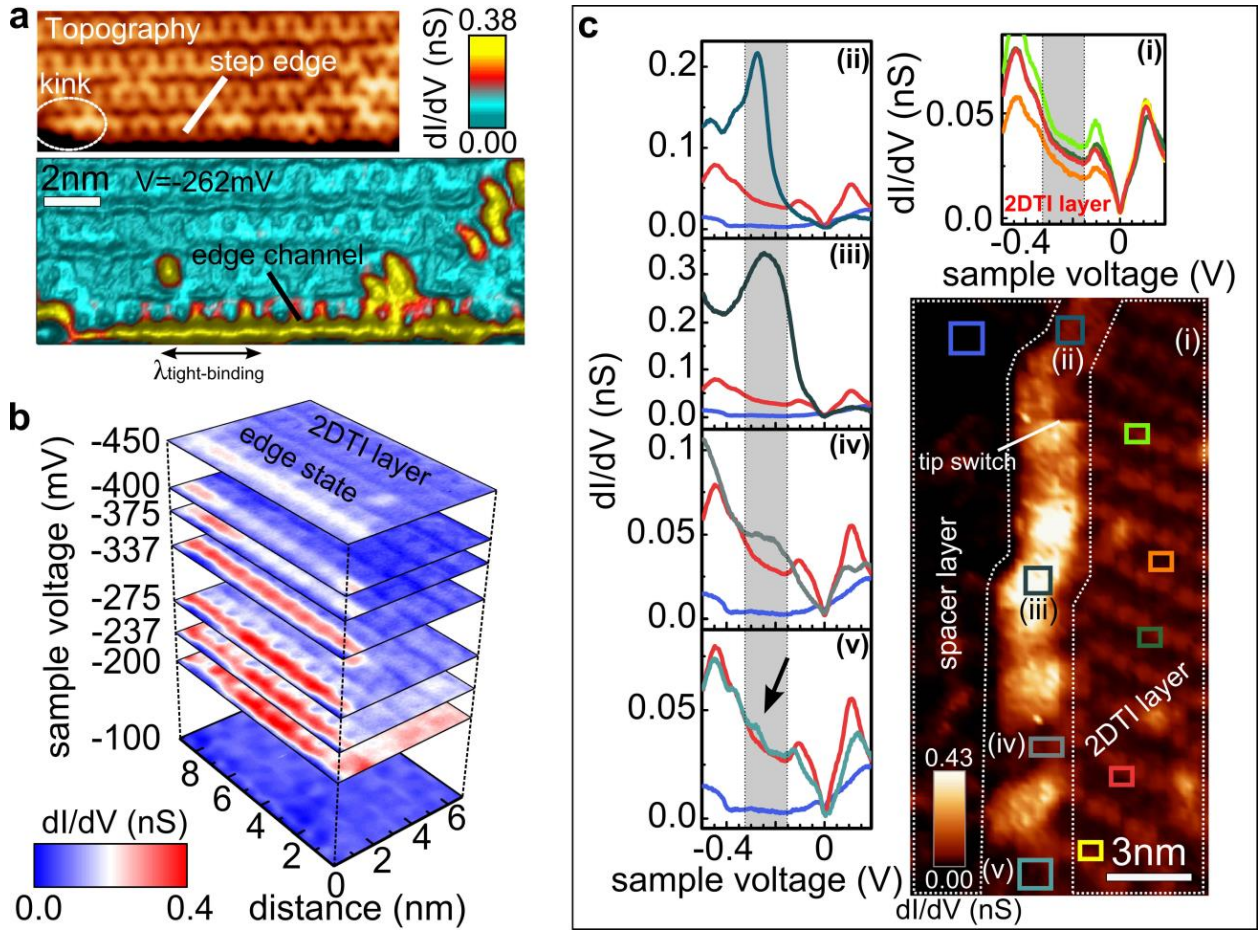

Fig.4

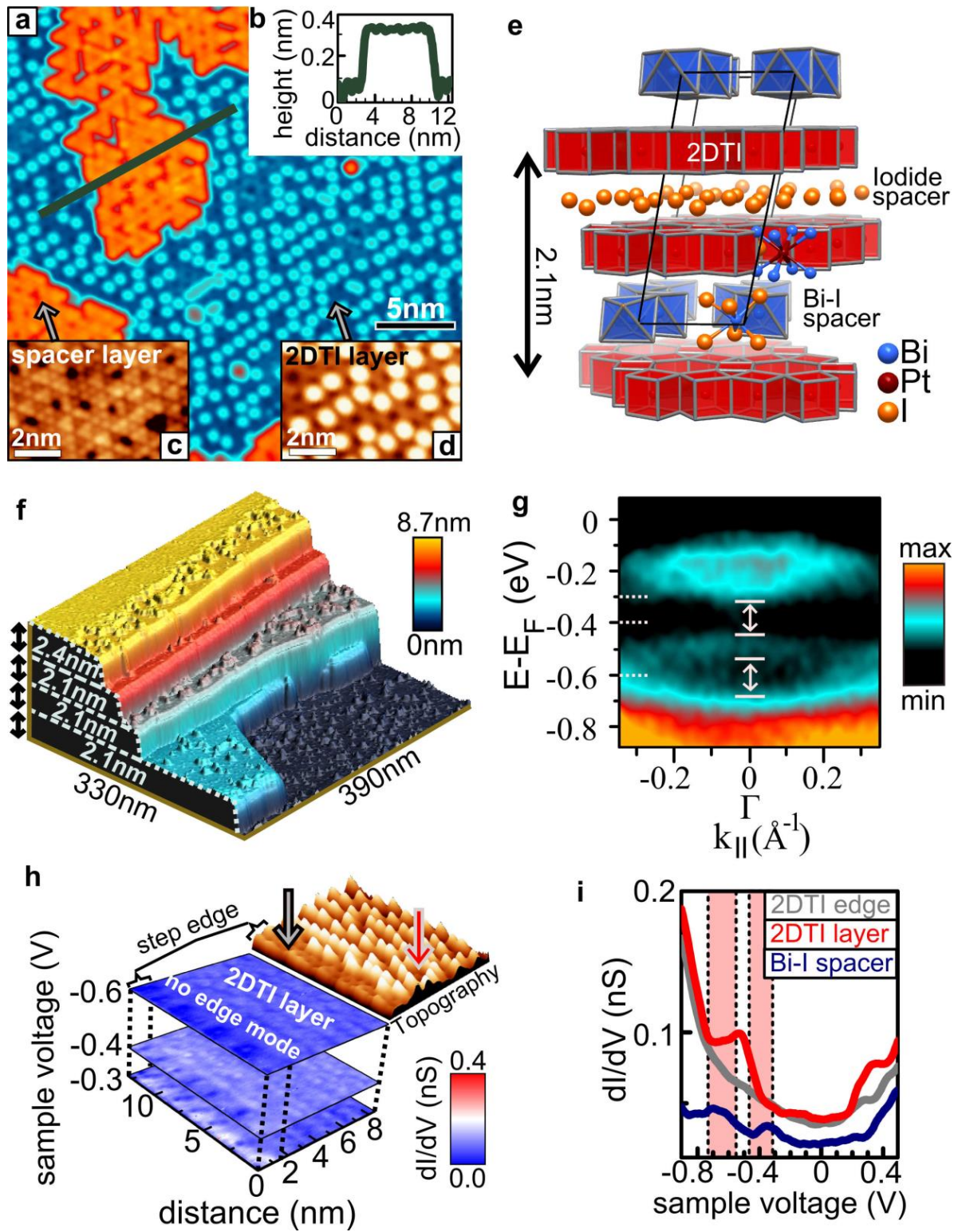

# Supplementary information for "Sub-nm wide electron channels protected by topology"

Christian Pauly[1], Bertold Rasche[2], Klaus Koepernik[3], Marcus Liebmann[1], Marco Pratzer[1], Manuel Richter[3], Jens Kellner[1], Markus Eschbach[4], Bernhard Kaufmann[1], Lukasz Plucinski[4], Claus M. Schneider[4], Michael Ruck[2,5], Jeroen van den Brink[3], and Markus Morgenstern[1]

[1]II. Institute of Physics B and JARA-FIT, RWTH Aachen University, D-52074 Aachen, Germany

[2]Department of Chemistry and Food Chemistry, TU Dresden, D-01062 Dresden, Germany

[3]Leibniz Institute for Solid State and Materials Research, IFW Dresden, P.O. box 270116, D-01171 Dresden, Germany and Dresden Center for Computational Science

[4]Peter Grünberg Institute PGI-6, Forschungszentrum Jülich, D-52425 Jülich, Germany

[5]Max Planck Institute for Chemical Physics of Solids, D-01187 Dresden, Germany

## S1: Continuity of the edge state

Figure S1a shows the topographic image of the step edge used for the *dI/dV* map of Fig. 3c in the main text. The *dI/dV* map itself is again displayed as Fig. S1b using a different color code. The intensity in the course of the edge state is reduced at positions of the step edge, where either a kink or two adsorbates are located (see dashed ellipses in Fig. S1c and S1d). The displacement of the *dI/dV* intensity towards the interior of the 2DTI layer indicates that the edge channel is pushed away from the step edge and simply moves around the obstacle[1]. The decrease in intensity shows that the edge state is additionally broadened in all three directions. In the case of the pair of adsorbates, we find some edge state intensity, if we place the tip on top of the adsorbates (spectrum (v) in Fig. 3c), revealing that the edge state at least partly channels below these adsorbates.

The diagonal stripe-like intensity fluctuation within the interior of the 2DTI layer has a different origin, as has been checked by measurements at different energies, which are compared to DFT. The stripes originate from the different coupling of the honeycombs of the 2DTI to the underlying stripe like pattern of the insulating spacer layer giving rise to a commensurability induced superstructure. Where cubes of the honeycomb are sitting on top of an octaeder of the spacer, the orbital weights of the bulk bands of the 2DTI is moved into the direction of the spacer and, thus, away from the surface. This leads to a shorter tip-surface distance, resulting in an increased background intensity at the energies of the 2DTI band gap and of the occupied bands (see Fig. 3c (i)) due to the finite tip LDOS at $E_\text{F,sample}$. The different origin can also be seen directly in the spectra. While the *dI/dV(V)* curves at the bright stripes (Fig. 3c (i)) are similarly increased at all negative voltages with respect to neighboring areas of the 2DTI layer, always showing a gap-like feature, the spectra on the edge (Fig. 3c (v)) exhibit an exclusive small peak around -0.3 V. Notice that the topologically protected edge states continuously evolve from the edge states known to be present at zig-zag edges of the honeycomb lattice[2], which, however, are prone to backscattering.

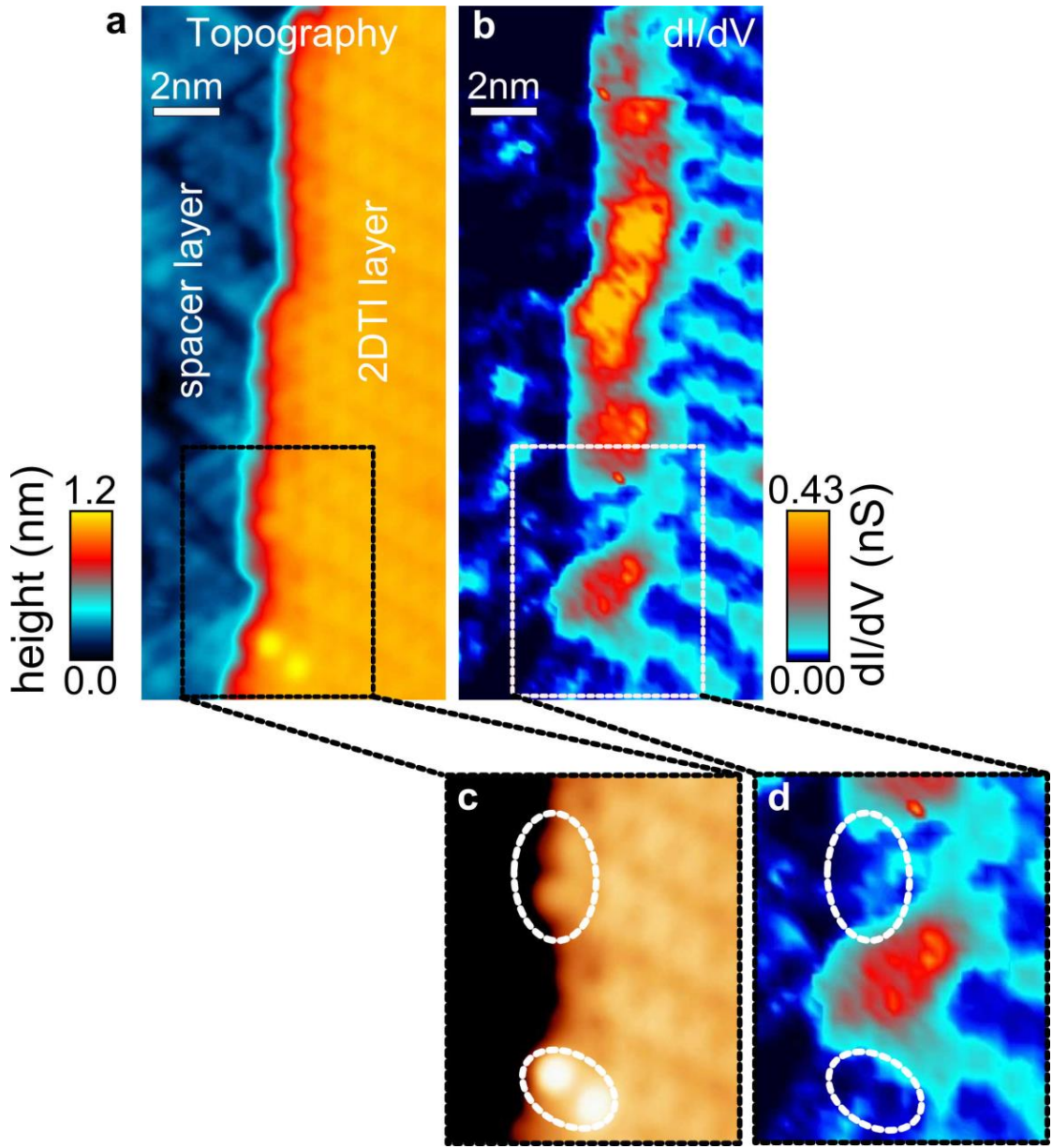

**Figure S1. Continuity of the edge state. a,** STM image of the step edge region of the 2DTI layer, which is also probed in Fig. 3c of the main text ($V$ = 0.8 V, $I$ = 80 pA). **b,** *dI/dV* image within the band gap of the 2DTI layer (averaged from $V$ = -180 mV to -350 mV, $I_{stab}$ = 80 pA, $V_{mod}$ = 4 mV) of the same area as (**a**) (Same as Fig. 3c of main text). **c, d,** Zoom into (**a**), (**b**) as marked by rectangles. Dashed ellipses highlight the positions of the kink and the two adsorbates.

## S2: DFT results and topological analysis on the electronic structure of $Bi_{13}Pt_3I_7$

Figure S2c shows the electronic band structure of a single 2DTI layer with the composition $[Bi_{12}Pt_3I]^{3+}$ as calculated by DFT. The displayed bands are numbered successively from the bottom to the top. Two band gaps marked in green are apparent around $E_F$. The parities of each band and the corresponding $Z_2$ indices valid in the energy region above the corresponding band are tabulated in Fig. S2e. $Z_2$ indices corresponding to band gaps are highlighted by dashed boxes. The first column of parities describes the Γ point, which is always a TRIM, while the other three points describe the remaining three TRIMs, i.e., the three different M points (see Fig. S2a). One observes that $Z_2$ gets non-trivial at band 5 below $E_F$ and gets trivial again at band 11 such that the marked band gaps around $E_F$ are topologically non-trivial containing edge states. Consequently, the $[Bi_{12}Pt_3I]^{3+}$ layer is a 2DTI according to LDA-PW92.

The exchange of parities between bands is quite complex. Multiple avoided crossings between the different valence bands, all dominated by Bi 6p orbitals from the 2DTI layer, can be conjectured from the individual band courses. We have cross-checked that these avoided crossings also cause a change of orbital character of the bands.

The same 2DTI properties with a similar complexity in parity exchange have been found for single 2DTI $[Bi_{12}Rh_3I]^{2+}$ layers. The three bands (12, 13 and 14) surrounding the non-trivial band gaps (Fig. S2g) look nearly identical to the bands 9, 10, and 11 of $[Bi_{12}Pt_3I]^{3+}$ except for a chemical shift by 0.3 eV upwards, which is caused by the different numbers of electrons within the 2DTI layers. Thus, the two materials probed in this study consist of very similar 2DTIs with honeycomb structure.

The DFT results for band structure, parities and $Z_2$ indices of the 3D semimetal $Bi_{13}Pt_3I_7$ are displayed in Fig. S2d and S2f. For the sake of simplicity, we artificially changed the arrangement of the BiI polyhedra within the spacer layer, which leaves the in-plane unit cell unchanged with respect to the original 2DTI layers. However, we checked that only small changes in the courses of the displayed bands are observed by performing DFT calculations of the real atomic arrangement as deduced from XRD. At least, for the bands 9 and 10 of the 2DTI, a doubling of the bands is apparent. It is marked by using the same band numbers in Fig. S2c and S2d with additional labels a and b in Fig. S2d. The band doubling originates from the doubling of the unit cell due to "dimerization". The pairs of bands exhibit exactly inverted parities at each TRIM indicating that they represent the bonding and anti-bonding linear combinations of the corresponding original bands of the two 2DTIs in the unit cell. Since the product of the parities at the TRIMs for each pair of such bands results in a minus sign, we find a trivial band topology for each pair of bands. Thus, we suggest that a simple doubling of the unit cell renders the weak TI trivial. A more rigorous analysis, however, has to justify this conjecture.

Figure S2h compares the calculated band structure with the ARPES results of the cleaved $Bi_{13}Pt_3I_7$. The Fermi level is shifted in order to account for possible surface doping effects. As in the case of $Bi_{14}Rh_3I_9$, the calculated bands have to be shifted down by about 0.3 eV indicating surface charging due to the cleavage process. The slightly smaller value compared to $Bi_{14}Rh_3I_9$ (~ 0.4 eV)[3] might be explained by the remaining iodide ions on the surface of the 2DTI layer in $Bi_{13}Pt_3I_7$ (Fig. 4a and 4d), which reduces the n-doping with respect to $Bi_{14}Rh_3I_9$. One band is obviously not visible in the ARPES data probably due to matrix element effects.

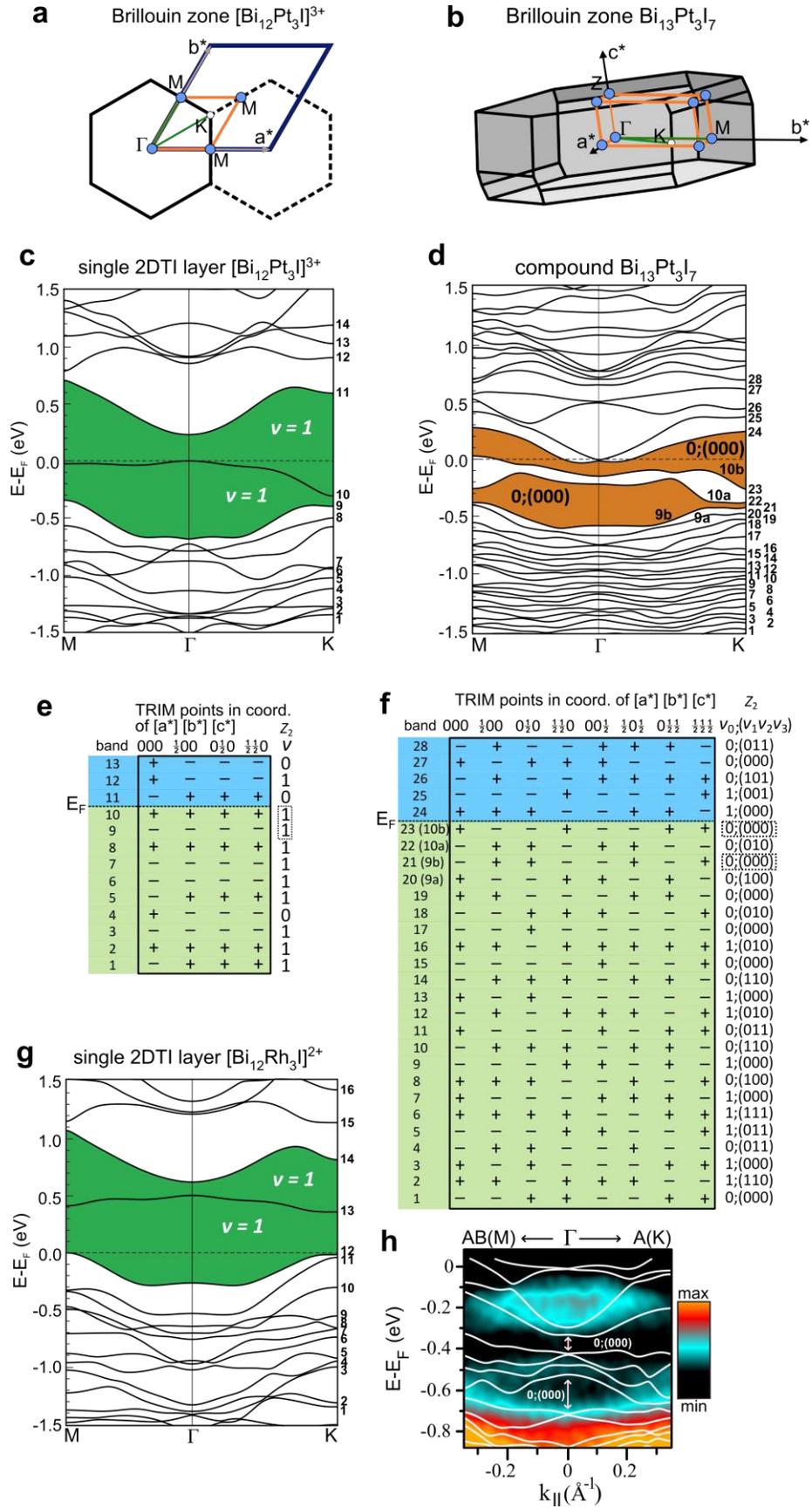

**Figure S2. Electronic structure of $Bi_{13}Pt_3I_7$ from DFT and ARPES: a,** Brillouin zone (black hexagons) of the 2DTI $[Bi_{12}Pt_3I]^{3+}$ with high symmetry points (Γ, K, M) marked. The orange diamond connects the TRIMs (marked by blue points) and the blue lines show the corresponding unit cell. The k-space directions of the displayed band structure in (**c**) are marked by green lines. Coordinate directions a* and b* in reciprocal space are additionally marked. **b,** Same as (**a**) but for the 3D material $Bi_{13}Pt_3I_7$. **c,** Fully relativistic DFT band structure of 2DTI layer $[Bi_{12}Pt_3I]^{3+}$ with numbered bands as used in (**e**). Green areas mark topological band gaps with calculated $Z_2$ indices marked. **d,** Same as (**c**) for the 3D material $Bi_{13}Pt_3I_7$ using a modified spacer layer leading to the same lateral unit cell as for $[Bi_{12}Pt_3I]^{3+}$. Pairs of bands are additionally labeled by the same numbers as in (**c**), but index a and b. Trivial band gaps are marked in orange with $Z_2$ indices marked (The area between band 23 and 24 is not a real gap within DFT, since band 24 at Γ and band 23 at M slightly overlap in energy). **e,** Table of parities of the different bands as numbered in (**c**) at the different TRIMs for $[Bi_{12}Pt_3I]^{3+}$. The $Z_2$ indices valid in the energy region above the corresponding band are shown on the right. $Z_2$ indices of band gaps are highlighted by dashed boxes. **f,** Same as (**e**) for the 3D material $Bi_{13}Pt_3I_7$. **g,** Same as (**c**) for the 2DTI $[Bi_{12}Rh_3I]^{2+}$. **h,** ARPES intensity plot of cleaved $Bi_{13}Pt_3I_7$ at $hv$ = 21.2 eV (same as Fig. 4g of the main text) with overlaid, calculated band structure (fully relativistic) using the experimental unit cell as deduced from XRD.

## S3: Carving edges into $Bi_{14}Rh_3I_9$

As described in the main text, the topological nature of the edge state and, in particular, its prohibited backscattering will cause electrons with opposite spins moving in opposite directions at each of these step edges and, thus, being protected against localization. If a step edge includes a 2DTI layer, topology guarantees the presence of an edge state. Thus, a step height above 1.3 nm is sufficient to necessarily include such an edge state. Eventually, tailored step edges could be connected to each other leading to quantum networks not suffering from backscattering and localization and to natural spin filters. To this end, we use the contact mode of the atomic force microscope (AFM) to scratch partially straight step edges into the surface.

Figure S3 shows an artwork representing the chemical symbols of the material, while Fig. 2d in the main text shows a network of AFM induced cuts leading to close-by edges with distances down to 25 nm. A further reduction of the distance between opposite step edges, short enough for electron tunneling, would lead to a beam splitter for electrons, which could come out of the splitter either left or right after travelling the parallel close-by step edges long enough.

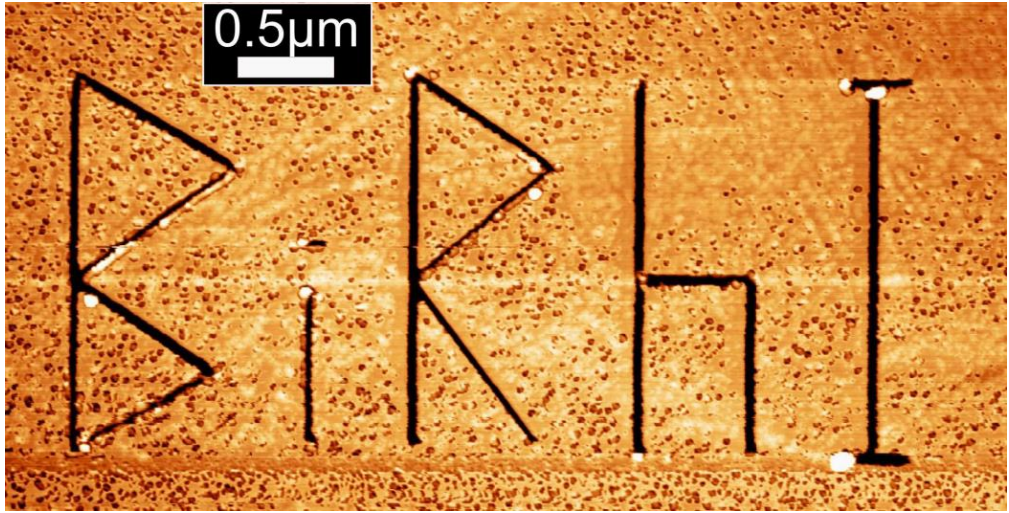

**Figure S3. Step edges scratched into Bi$_{14}$Rh$_3$I$_9$ by atomic force microscopy (AFM).** AFM image of Bi$_{14}$Rh$_3$I$_9$ surface with letters BiRhI scratched into the surface by a carbon coated silicon cantilever in AFM contact mode at ambient conditions (contact force during scratching: $F = 10^{-6}$ N). Average depth of the cuts: ~ 15 nm. The AFM images have been recorded in the tapping mode using the same carbon coated silicon cantilever (resonance frequency 275.1 kHz, force constant 43 N/m, oscillation amplitude 30 nm, set point 70%, velocity 2 μm/s).